\documentclass[final]{vgtc}                 

\graphicspath{{figures/}{pictures/}{images/}{./}} 

\usepackage{times}                     

\usepackage{tabu}                      
\usepackage{booktabs}                  
\usepackage{lipsum}                    
\usepackage{mwe}                       

\usepackage{mathptmx}                  
\usepackage{amsmath}
\usepackage{amssymb}

\usepackage{tabularx}
\usepackage{array}
\def\our{DomeGS}
\usepackage{todonotes}

\onlineid{1955}

\vgtccategory{Research}

\vgtcinsertpkg

\title{\our{}: Perceptually Consistent Background Completion for Unbounded 3D Gaussian Splatting in Virtual Reality}

\author{Anastasiya Pechko\thanks{e-mail: anastasiya.pechko@doctoral.uj.edu.pl, Doctoral School in the Exact Sciences, Jagiellonian University}\\ %
    \scriptsize Jagiellonian University %
\and Agata Szymańska \thanks{e-mail:  agata.szymanska@doctoral.uj.edu.pl, Doctoral School in the Social Sciences, Jagiellonian University,
Krakow, Poland}\\ %
     \scriptsize Jagiellonian University %
\and Piotr Borycki \thanks{e-mail: piotr.borycki@doctoral.uj.edu.pl, Doctoral School in the Exact Sciences, Jagiellonian University}\\ %
     \parbox{1.4in}{\scriptsize \centering  Jagiellonian University\\ Ideas Research Institute} %
\and Joanna Szelewska \thanks{e-mail:  joanna.szelewska@doctoral.uj.edu.pl, Doctoral School in the Exact Sciences, Jagiellonian University}\\ %
     \parbox{1.4in}{\scriptsize \centering  Jagiellonian University\\ Ideas Research Institute} %
\and Michał Mnich \thanks{e-mail: michal.mnich@uj.edu.pl}\\ %
     \scriptsize Jagiellonian University %
\and Slawomir Konrad Tadeja \thanks{e-mail: skt40@cam.ac.uk}\\ %
     \scriptsize University of Cambridge %
\and Przemysław Spurek\thanks{e-mail: przemyslaw.spurek@uj.edu.pl}\\ %
     \parbox{1.4in}{\scriptsize \centering  Jagiellonian University\\ Ideas Research Institute}
}
\teaser{
  \centering
  \includegraphics[width=\linewidth]{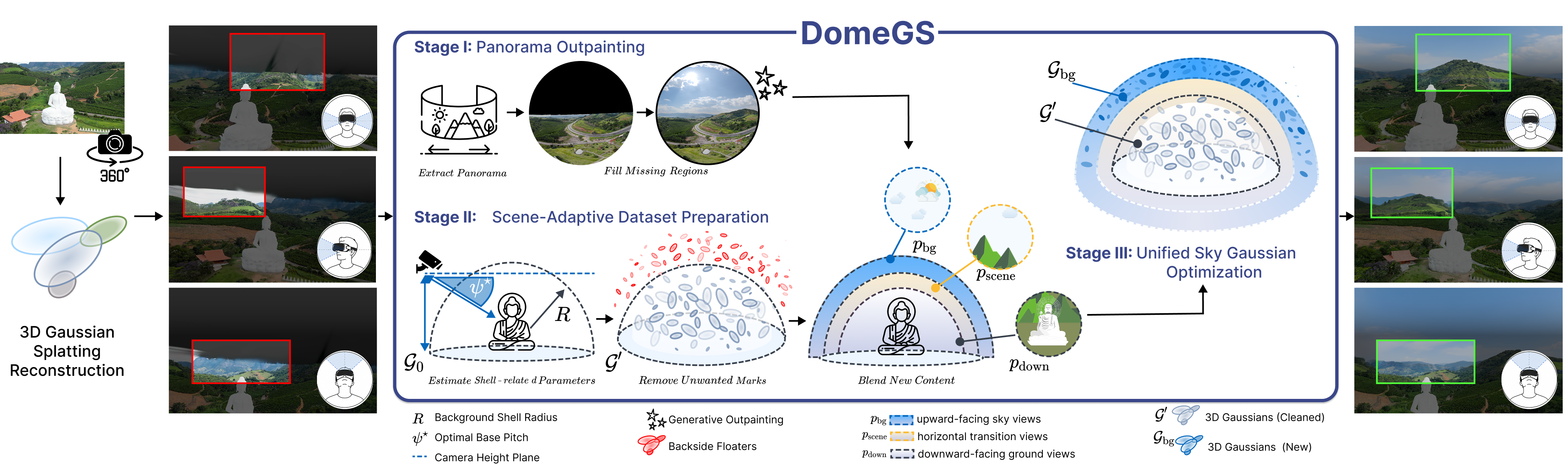}
  \caption{
    We present \our{} a method for filling missing background in 3D Gaussian Splatting scenes. It is directly incorporated with the scene by utilizing outpainted panorama view, followed by scene-adaptive dataset preparation, which expands the set of training views to incorporate previously missing background. Finally, the scene is fine-tuned according to both original and added views to achieve a unified Gaussian Splatting scene. This design encourages preserving depth and realism, enabling immersive VR environment creation.
}
  \label{fig:teaser}
}

\abstract{
The growing need for high-quality 3D content remains among the most prevalent challenges in \textit{virtual reality} (VR). Recently, novel view synthesis techniques have enabled rapid content creation, with the \textit{3D Gaussian Splatting} (3DGS) emerging as one of the most promising methods for generating high-fidelity 3D scenes. However, using 3DGS to reconstruct unbounded outdoor scenes can leave horizons incomplete due to missing data or a lack of distant geometric constraints. Such incompleteness can break immersion, compromising VR's core characteristic. A typical workaround in traditional VR systems is a textured cube (\textit{skybox}) or sphere (\textit{skydome}) surrounding the environment to create the illusion of a distant background. However, these approaches struggle to blend seamlessly with the reconstructed 3DGS geometry. To address these issues, we introduce the \our{} framework \footnote{Code: \url{https://github.com/Anastasiya999/DomeGS}}, which incorporates the skydome concept not as a flat background shell, but as a part of the reconstructed Gaussian Splatting 3D scene. By integrating the horizon and sky into the 3DGS representation, our method provides a unified, scene-consistent environment bridging foregrounds and backgrounds for greater realism. Experimental evaluation with $N=36$ users demonstrates that \our{} significantly increases spatial presence ($p=0.042$) and realism ($p=0.042$) compared to the skybox baseline.
} 

\keywords{Virtual reality, VR, skybox, skydome, Gaussian Splatting, 3DGS.}

\begin{document}

\maketitle

\section{Introduction}
Realistic, highly immersive experiences are a core requirement in spatial interfaces such as \textit{virtual reality} (VR) \cite{slater2009place, cummings2016immersive} and a primary objective within the spatial computing community \cite{huang_personalized_2025, yuan2025immersegen}. A key factor shaping these experiences is the high-quality 3D content that populates fully virtual environments \cite{tadeja_2020}. Recently, \textit{3D Gaussian Splatting} (3DGS) \cite{kerbl3Dgaussians} has emerged as a promising method of scene representation and rendering for interactive VR systems, providing high-fidelity 3D reconstruction while also enabling complex physics-aware interactions \cite{jiang2024vr, pechko2025gs}. 3DGS is a real-time view synthesis method that offers high visual quality at frame rates feasible for head-mounted displays (HMDs). 

3DGS works best when reconstructing near-field, bounded scenes with compelling depth cues. However, when we attempt to reconstruct unbounded outdoor environments, 3DGS reveals a critical constraint caused by sparseness or entirely missing far-field detail in the capture data (e.g. skies, distant facades, and the horizon). Without sufficient multi-view parallax and geometric landmarks, the 3DGS optimization process lacks the details needed to accurately triangulate depth in these regions \cite{barron2022mip}. As a result, standard 3DGS models fail to reconstruct plausible geometry in the distance \cite{pintani2026envmap, wang2025look}. In immersive VR, this failure can negatively affect immersion, as users may encounter jagged boundaries, floating artifacts, or gaps in the virtual environment that break the illusion of a continuous unbounded outdoor scene \cite{xiong2023sparsegs, yuan2025robust} (see Fig.~\ref{fig:teaser}).

\begin{figure}[t]
  \centering
  \includegraphics[width=\linewidth]{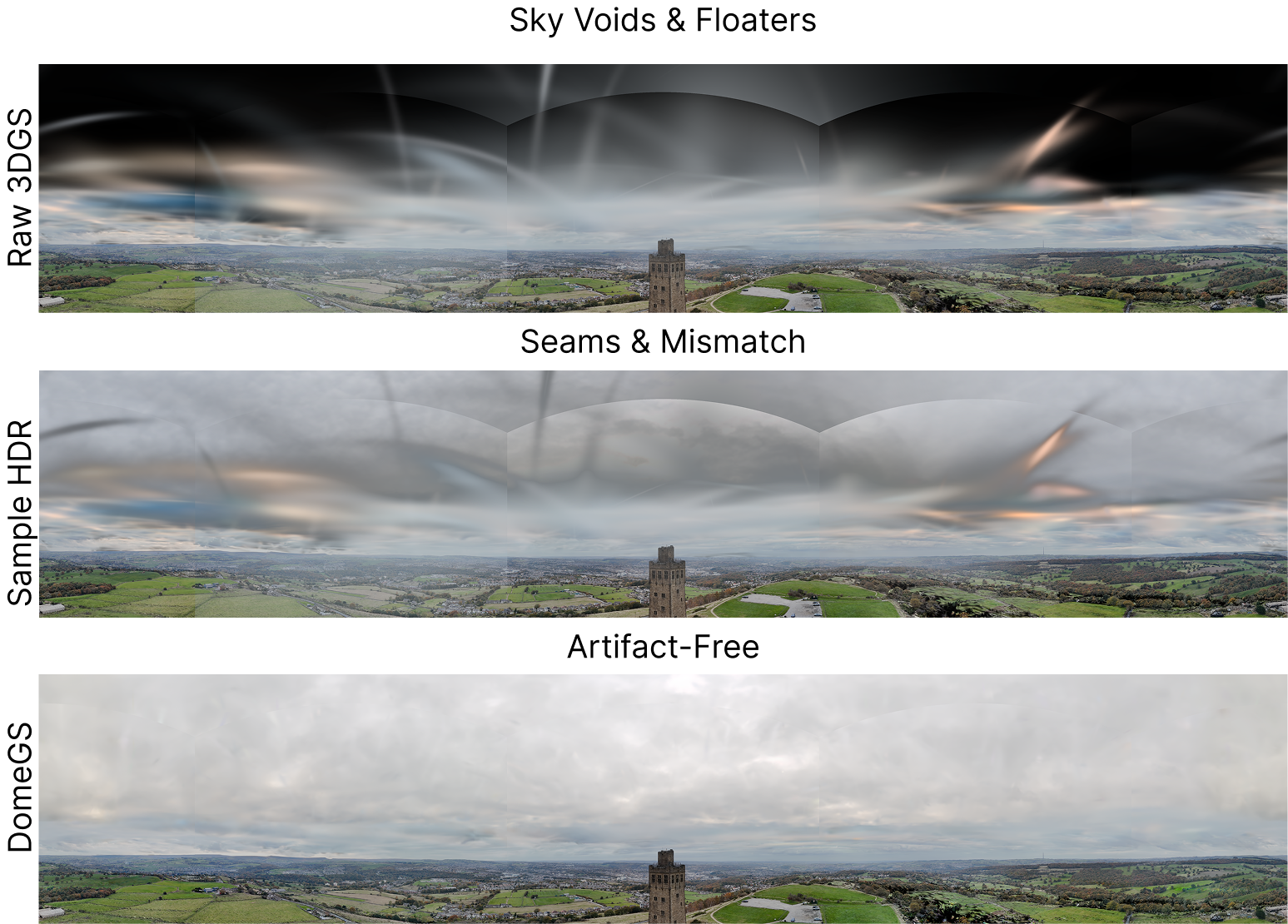}
\caption{3DGS scene integration in game engines, illustrated using Unity. Top row: Raw 3DGS, rendered without any background completion, exhibiting missing sky regions and boundary floaters. Middle row: Naive integration with a cubemap skybox fills the missing background but introduces visible seams and illumination mismatch. Bottom row: Our method achieves seamless background integration without visible artifacts.
}
\label{fig:scenes_and_problems}
\vspace{-4mm}
\end{figure}

A common workaround to mask these missing regions, natively supported in game engines such as Unity\footnote{Unity 6.1 User Manual, \url{https://docs.unity3d.com/6000.1/Documentation/Manual/skyboxes-using.html}}, is to project a panoramic image onto an infinitely distant skybox. Here, recent advancements have made significant strides in unbounded 3D scene generation~\cite{chai2023persistent, chen2023scenedreamer} through development of sophisticated diffusion models capable of synthesizing high-quality $360^\circ$ panoramas~\cite{ kalischek2025cubediff, wu2024panodiffusion, ye2024diffpano}, and extended them to layered 3D representations~\cite{yang2025layerpano3d, zhou2024dreamscene360}. While these completion methods can successfully synthesize missing backgrounds, modular frameworks facilitate their integration \cite{ye2024gaustudio}, treating the background as an infinite cubemap. However, directly compositing such a pre-existing map with the reconstructed scene can cause color or illumination discrepancies and result in inconsistent, implausible integration between the reconstructed scene and the panoramic background, visually noticeable between the near-field Gaussians and the far-field backdrop (see Fig.~\ref{fig:scenes_and_problems}). 

To address this, we introduce \our{}, a novel framework for perceptually consistent background completion for 3DGS scene reconstruction. Rather than relegating the far-field to a post-processed 2D backdrop, our method constructs a scene with a background inspired by skydome-like approaches (i.e. an alternative to a skybox solution based on a textured sphere instead of a cube to surround the virtual environment) directly within the Gaussian representation. We first extract a partial panorama from the near-field scene and complete the missing distant background regions using a $360^\circ$-specialized generative model. Crucially, we then use this completed map to supervise a secondary optimization stage, where we seed and refine new Gaussians on a finite-radius distant sphere. This unification of the near-field scene and the far-field skydome into a single 3DGS asset offered by our \our{} ensures that the entire environment responds cohesively to user movement. The result is a geometrically continuous background plausibly integrated with the 3DGS scene. Furthermore, we experimentally compared the \our{} method with the skybox baseline and showed significant increases in sense of presence ($p=0.042$) and realism ($p=0.042$). In summary, our main contributions are threefold:
\begin{enumerate}
    \item We offer a unified pipeline for extending unbounded 3DGS scenes specifically designed to preserve depth cues and realism in an immersive VR environment.
    
    \item We effectively integrate $360^\circ$-aware generative outpainting into a Gaussian-based skydome, achieving photometric alignment and coherent background completion without degrading the primary reconstructed scene.
    
    \item Results of a comprehensive within-subjects user study demonstrating that our \our{} methods significantly enhance 3DGS-based background consistency under motion, spatial presence, and experienced realism compared to the classical infinite skybox baseline while eliciting comparable behavior and comfort.
\end{enumerate}

\section{Related works}\label{sec:related_work}
Here, we first reviewed classical approaches to representing distant backgrounds in VR, such as \textit{skybox} and \textit{skydome}. Next, we reviewed generative methods for 3D environment completion as well as their current limitations. Finally, we specifically reviewed limited approaches for completing 3DGS reconstructions. 

\subsection{Distant Background Representations for VR} 
A standard way to create the illusion of distant backgrounds in 3D environments is the \textit{skybox} approach, which maps panoramic imagery onto an effectively infinite cubemap \cite{greene1986environment}, as a high-dynamic-range image (HDRI) to encode both appearance and illumination for image-based lighting. In conventional pipelines, such a decoupling strategy, where the skybox surrounds explicit mesh geometry representing foreground elements, allows environmental conditions to be adjusted simply by swapping the HDRI, without touching the underlying meshes. An alternative approach, namely \emph{skydome}, represents the environment on a finite spherical or hemispherical surface, introducing depth-dependent motion parallax under viewer translation. Unlike the skybox, it must be manually constructed and integrated as a custom mesh and shader\footnote{Skydome tutorial: \url{https://www.rastertek.com/tertut10.html}.}, making it unsuitable as a baseline without dedicated implementation. Although less common as a default environment representation, this finite spatial extent is relevant to VR, where motion parallax provides important cues for depth, self-motion, and scene scale~\cite{beall1995absolute, howard2002seeing}.

\subsection{Generative Environment Completion} 
Generative methods have recently enabled the synthesis of environment content beyond the observed regions, supporting unbounded scene generation, novel-view extrapolation \cite{chai2023persistent, chen2023scenedreamer}, as well as full $360^\circ$ panorama generation~\cite{kalischek2025cubediff, wu2024panodiffusion, ye2024diffpano}. In a conventional pipeline, a pre-existing $360^\circ$ panoramic file, such as a high dynamic range image (HDRI), can be freely replaced or adjusted as the foreground geometry and its appearance are controlled independently. This differs for scenes reconstructed from captured images, such as a 3DGS-based scene, because the environment's appearance and illumination are implicitly baked into the reconstruction, so an independently chosen HDRI may introduce inconsistent colors, lighting, or visual cues at the foreground-background boundary. Therefore, the missing environment must be synthesized consistently with the 3D scene rather than treated as an independent component.

Recent methods have further extended generative environment synthesis to explorable 3D representations by converting generated panoramas into 3D Gaussian representations~\cite{yang2025layerpano3d, zhou2024dreamscene360}. However, these approaches generate new environments or scenes rather than completing the unobserved surroundings of an existing 3DGS reconstruction under explicit scene-consistency constraints. To the best of our knowledge, this setting remains unexplored. Our work addresses this gap by generating novel views conditioned on the captured scene and reconstructing them as a finite Gaussian skydome around the existing 3DGS representation, extending the scene with spatially coherent background content while preserving depth-dependent motion parallax.

\subsection{3DGS for Unbounded Scene Reconstruction} 
3D Gaussian Splatting (3DGS) \cite{kerbl3Dgaussians} combines high-fidelity view synthesis with real-time rendering for high-quality, thus attracting increasing interest from the domain of spatial interfaces \cite{jiang2024vr} as it can reconstruct indoor and outdoor scenes, as well as support other applications \cite{li2025radiance, schieber2025semantics}. However, its structure-from-motion (SfM)-based initialization provides limited support for plausibly reconstructing distant regions of the environment, such as the skies or horizon, especially since their details are often sparse or missing from the input imagery data. This is particularly challenging for using 3DGS to reconstruct unbounded outdoor scenes, where nearby scene content may be well represented while distant environmental regions remain unobserved.

Existing methods address this by removing or separately modeling such
regions, including sky masking in WildGaussians~\cite{kulhanek2024wildgaussians}
and a finite spherical Gaussian shell in EnvMap-GS~\cite{pintani2026envmap}.
EnvMap-GS fits that shell to observed background radiance and can bake it
into an environment map; it does not generate missing sky, and baking
again yields an infinite, no-parallax backdrop. 

However, these approaches do not address synthesizing environmental content that is entirely absent from the input imagery. Moreover, their evaluation typically considers static viewpoints rather than movement in immersive environments, where continuous viewpoint changes can reveal inconsistencies in distant scene content and horizon cues.



\section{Methodology}
\label{sec:method}

The goal of our method is to create an impressive environment featuring a reconstructed scene and additional outpainting background (see  Fig.~\ref{fig:teaser}). This section first introduces the necessary preliminaries and then presents our proposed pipeline, which consists of three main stages. (I) \textit{Panorama Extraction and Outpainting} generates a seamless equirectangular environment map from the existing scene; (II) \textit{Scene-Adaptive Dataset Preparation} converts that 2D panorama into a geometrically grounded, multi-view supervision dataset; and (III) \textit{Unified Sky Gaussian Optimization} fuses this dataset with the original scene into a single 3DGS representation. 

\subsection{Preliminaries and Problem Formulation}
\label{sec:formulation}
We develop our method on top of 3DGS~\cite{kerbl3Dgaussians}, which represents a 3D environment as an unstructured collection of explicit 3D Gaussians $\mathcal{G} = \{g_i\}_{i=1}^N$. Each primitive $g_i$ is parameterized by a tuple $(\mu_i, \Sigma_i, \alpha_i, c_i)$, where $\mu_i \in \mathbb{R}^3$ is its center position, $\Sigma_i \in \mathbb{R}^{3\times 3}$ is a 3D covariance matrix (factorized into scaling $S_i$ and rotation quaternion $R_i$), $\alpha_i \in [0, 1]$ is the opacity, and $c_i$ denotes view-dependent spherical harmonics (SH) coefficients. Given a 6DoF camera pose $\mathbf{T} = [\mathbf{R} \mid \mathbf{t}] \in \mathrm{SE}(3)$ representing an HMD  eye viewpoint, primitives are projected onto the image plane and alpha-blended via a tile-based rasterizer to render novel views in real time. 

Achieving immersive experience in VR relies on unrestricted 6DoF navigation. However, an initial scene $\mathcal{G}_0 = \{g_i\}_{i=1}^{N_0}$ reconstructed from a collection of training views leaves peripheral regions unmodeled. Standard, naive background completion fails because immersion depends on two conditions:

(1) \textbf{Radiometric Coherence:} The created background must seamlessly match the baked illumination, tone, and color balance of $\mathcal{G}_0$, preventing horizon boundary artifacts.

(2) \textbf{Kinematic Parallax:} Under translational head movement $t_{\text{HMD}}$, the background must preserve realistic, depth-consistent motion parallax rather than being static, as in traditional skyboxes.


To achieve these requirements, we formulate our objective as expanding $\mathcal{G}_0$ with a background Gaussian collection $\mathcal{G}_{\text{bg}} = \{g_j\}_{j=1}^{N_{\text{bg}}}$ into a unified scene:
\(
    \mathcal{G} = \mathcal{G}_0 \cup \mathcal{G}_{\text{bg}}
\)
where $\mathcal{G}_{\text{bg}}$ is anchored on a distant, finite-radius spherical shell, enabling joint rasterization with $\mathcal{G}_0$ under arbitrary HMD poses.



\subsection{Stage I: Panorama Extraction and Outpainting}
\label{sec:stage1}

\begin{figure}[t]
  \centering
  \includegraphics[width=\linewidth ]{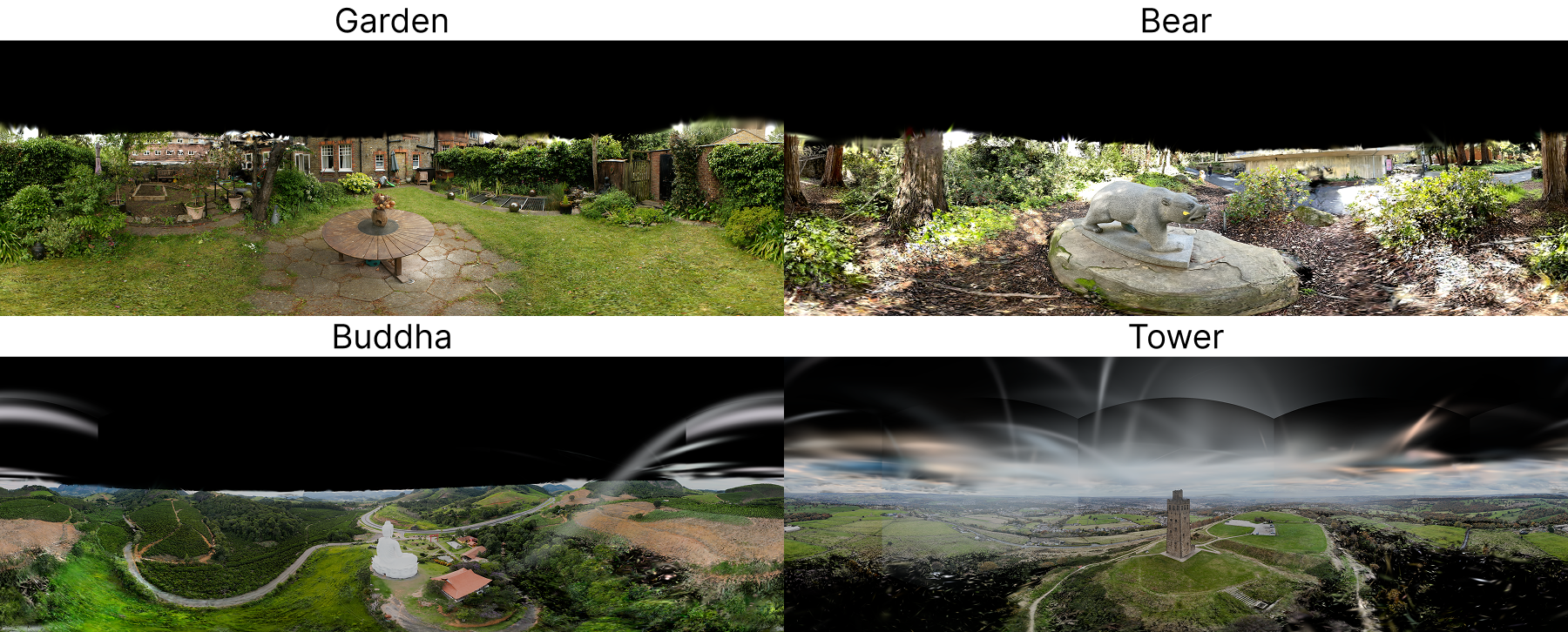}
\caption{ Experimental scenes used in our study, rendered as equirectangular projections from $\mathcal{G}_{\mathrm{pano}}$. The four outdoor unbounded 3DGS reconstructions are divided into two scene types: \emph{Close-range} (top: Garden, Bear) and \emph{Panoramic} (bottom:  Buddha,  Tower).}
  \label{fig:scenes}
  \vspace{-0.1cm}
\end{figure}

To initialize background completion we place a virtual panorama camera at the trajectory centroid $\mathbf{e} = \frac{1}{K} \sum_{k=1}^{K} \mathbf{t}_k$ and exclude Gaussians intersecting an exclusion sphere of radius $r$ around $\mathbf{e}$:
\begin{equation}
    \mathcal{G}_{\text{pano}} = \left\{\, g_i \in \mathcal{G}_0 \;\middle|\; \|\mu_i - \mathbf{e}\|_2 > r + s_i^{\max} \,\right\},
\end{equation}
where $s_i^{\max}$ is the maximum scaling factor of primitive $i$. We then render an equirectangular projection image $I_0$ from $\mathbf{e}$ using $\mathcal{G}_{\text{pano}}$~(Fig.~\ref{fig:scenes})\ and construct a binary completion mask $M$, where $M=1$ indicates unobserved sky and background voids. 

To complete the missing environment without boundary artifacts, we employ a two-stage generative process:

(1) \textbf{Global Outpainting:} A $360^\circ$-aware latent diffusion model~\cite{nomadoor_flux2_klein_360_erp_2026} is conditioned on $I_0$ and $M$ to fill missing background, generating an initial panorama $I^{(1)}$.

(2) \textbf{Wrap-Seam Harmonization:} To eliminate boundary discontinuities at the $360^\circ$ wrap seam, we cyclically shift $I^{(1)}$ by $180^\circ$ to recenter the panorama. A vertical strip mask $M_{\text{seam}}$ centered at the new boundary is inpainted with a secondary diffusion pass, yielding a seamless panorama s$I^{(2)}$.

\subsection{Stage II: Scene-Adaptive Dataset Preparation}
\label{sec:stage2}


\noindent\textbf{Geometric Parameter Estimation.}
To properly incorporate the 2D completed panorama $I^{(2)}$ within 3D space, we dynamically calculate two geometric parameters:

(1) \textbf{Background Shell Radius ($R$):} To ensure the skydome encompasses all foreground geometry without intersecting near-field objects, we estimate the shell radius from the distribution of visible scene depths rendered from centroid $\mathbf{e}$:
    \begin{equation}
        R = m \cdot \mathrm{Quantile}_q\!\left(\{\|\mu_i - \mathbf{e}\|_2 : i \in V_{\text{cube}}\}\right),
    \end{equation}
    where $V_{\text{cube}}$ denotes Gaussians visible across a 6-face cubemap, with quantile $q = 98\%$ and scale factor $m = 1$.

(2) \textbf{Optimal Base Pitch ($\psi^\star$):} Casual captures often exhibit an asymmetric vertical camera distribution. We determine an optimal base camera pitch $\psi^\star$ by evaluating probe cameras relative to the world up-vector $\mathbf{u}_{\text{up}}$, maximizing visible scene density while penalizing unobserved sky voids:
    \begin{equation}
        \psi^\star = \arg\max_\psi \left( \frac{|\mathcal{G}(\psi) \cap \mathcal{G}_{fg}|}{|\mathcal{G}_{fg}|} - \lambda_{\psi} \frac{|\mathcal{G}(\psi) \setminus \mathcal{G}_{fg}|}{|\mathcal{G}(\psi)|} \right),
    \end{equation}
    where $\mathcal{G}_{fg}$ is the set of confident foreground Gaussians with Gaussian opacity ($\alpha_i$) greater than or equal to a fixed threshold ($\tau_\alpha$), $\lambda_{\psi}=0.3$ is a hyperparameter, and $\mathcal{G}(\psi)$ is the set of primitives visible at pitch $\psi$.


In addition, 3DGS models trained on bounded trajectories frequently accumulate degenerate, low-opacity Gaussians outside the capture frustum. We must remove these unconstrained primitives to not corrupt the newly synthesized background. We cast a set of ``look-back'' probe cameras yaw-flipped by $180^\circ$ and pitched away from the primary scene ($\psi_{lb}$). Primitives visible in these views, $\mathcal{G}_{lb}$, are identified as backside floaters and pruned via
    $\mathcal{G}' = \mathcal{G}_0 \setminus \mathcal{G}_{lb}$. Full derivations of $\psi^\star$ and $\psi_{\mathrm{lb}}$, along with the fixed threshold $\tau_\alpha$ and search settings, are provided in the supplementary material.

With the cleaned scene $\mathcal{G}'$ and geometry parameters $R,\psi^\star$, we synthesize a multi-view dataset $I_{\text{sup}}$ across stratified camera rings at varying vertical pitch angles: downward ground views ($p_{\text{down}}$), horizontal transition views ($p_{\text{scene}}$), and upward sky views ($p_{\text{bg}}$). These bands serve distinct functional roles and are assigned role-dependent supervision targets:
\begin{equation}
I_{\text{sup}}(\mathbf{T}) =
\begin{cases}
I_{\mathcal{G}'}(\mathbf{T}),
& p = p_{\text{down}}, \\[2pt]
I_{\mathcal{G}'}(\mathbf{T})
+ I_{\text{bg}}(\mathbf{T}) \odot
\big(1-\alpha(\mathbf{T})\big),
& p \in \{p_{\text{scene}},p_{\text{bg}}\},
\end{cases}
\label{eq:supervision}
\end{equation}
where $I_{\mathcal{G}'}(\mathbf{T})$ and $\alpha(\mathbf{T}) \in [0, 1]$ are the rendered RGB image using camera $\mathbf{T}$ and accumulated 2D opacity from $\mathcal{G}'$ against a black background. The background image $I_{\text{bg}}(\mathbf{T})$ is obtained by casting camera rays from $\mathbf{T}$, computing their intersection with the bounding sphere of radius $R$ and sampling the completed panorama $I^{(2)}$.


\begin{figure}[t]
  \centering
  \includegraphics[width=\linewidth]{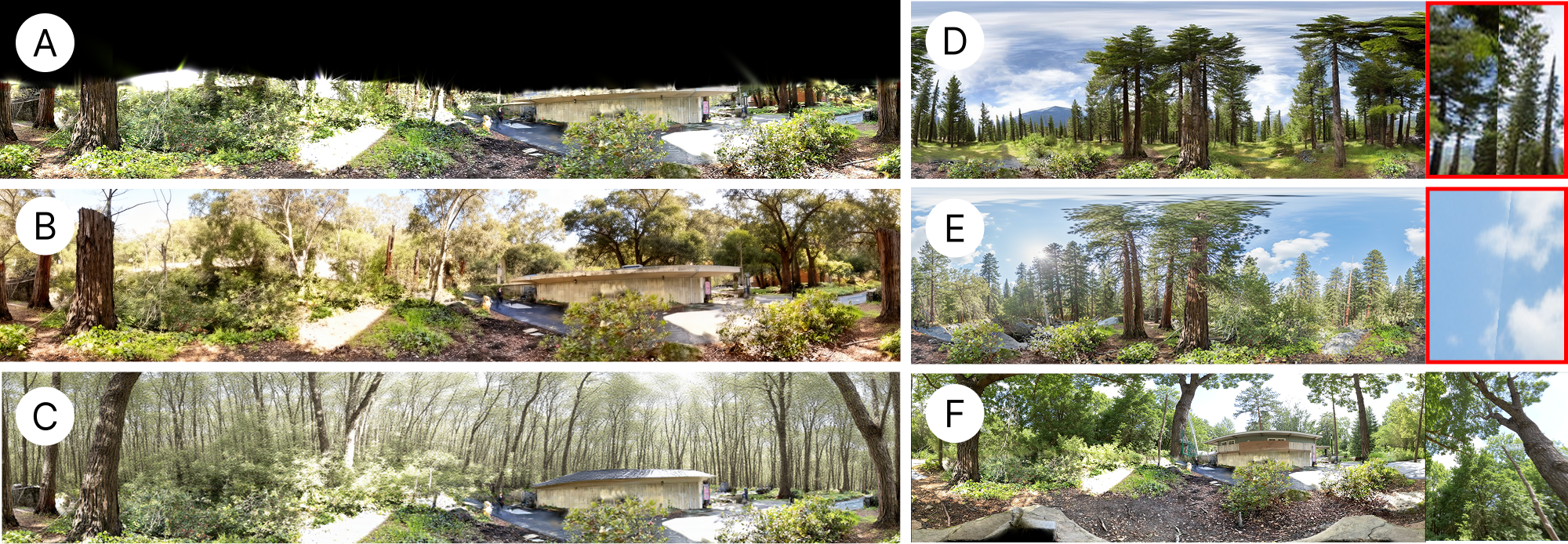}
  \caption{Comparison of generators for completing the same masked panorama: (A) initial panorama with missing regions; (B) FLUX.2-dev, with a color mismatch; (C) FLUX.1-Fill-dev, with repeated structures; (D) \texttt{512-inpainting-ema}, with discontinuity at the horizon; (E) Nano Banana, with a visible sky seam; and (F) FLUX.2-klein-base-9B and 360 ERP LoRA, with better color consistency and reduced seam artifacts, selected for the VR conditions.}
  \label{fig:ablation-gen}
  \vspace{-0.5cm}
\end{figure}

This role-dependent formulation resolves boundary seams by treating foreground silhouettes and distant background content through physically grounded alpha compositing. Consequently, it guides newly optimized background Gaussians to populate empty regions behind foreground geometry, eliminating boundary clipping without explicit mesh segmentation or manual refinement.


\subsection{Stage III: Unified Sky Gaussian Optimization}
\label{sec:stage3}

The optimization stage initializes from the floater-cleaned initial scene $\mathcal{G}'$ and metadata $(\mathbf{e}, R)$. To preserve captured foreground structures while integrating background into the scene, we fix Gaussians within radius $r_f = \max_k \|\mathbf{t}_k - \mathbf{e}\|_2$ throughout optimization. Primitives outside $r_f$ remain trainable and protected from pruning, size-based culling, and opacity resets. To represent the background, we initialize a set of seed Gaussians $\mathcal{G}_{\text{seed}}$ distributed across an upper-hemispherical shell centered at $\mathbf{e}$:
\begin{equation}
    \boldsymbol{\mu}_j =
    \mathbf{e} + r_j \mathbf{d}_j,
    \quad
    \mathbf{d}_j \sim \mathrm{Unif}(\mathbb{S}^2_+),
    \quad
    r_j \sim \mathrm{Unif}([\gamma R, R]),
\end{equation}
where $\mathbb{S}^2_+ = \{\mathbf{d}\in \mathbb{S}^2 \mid \mathbf{d}\cdot\mathbf{u}_\text{up}\ge 0\}$ denotes the unit upper hemisphere in the world coordinates, and $\gamma = 0.99$ defines the shell thickness across all scenes. These Gaussians are initialized with scales adapted to the local sampling density, then optimized together with the original scene.

At each training step, we sample a synthetic camera pose $\mathbf{T}$ alongside its supervision target $I_{\text{sup}}$ (Eq.~\ref{eq:supervision}), render the current scene $\hat{I}_{\mathcal{G}}(\mathbf{T})$, and minimize a joint photometric and geometric objective:
\begin{equation}
    \mathcal{L} = \mathcal{L}_{\text{photo}} + \lambda_{\text{shell}} \mathcal{L}_{\text{shell}} + \lambda_{\text{plan}} \mathcal{L}_{\text{plan}}.
    \label{eq:total_loss}
\end{equation}
The photometric loss $\mathcal{L}_{\text{photo}}$ combines an $\mathcal{L}_1$ color difference with structural dissimilarity (D-SSIM)~\cite{kerbl3Dgaussians}:
\begin{equation}
    \mathcal{L}_{\text{photo}} = (1 - \lambda_{\text{ds}})\, \|\hat{I}_{\mathcal{G}} - I_{\text{sup}}\|_1 + \lambda_{\text{ds}}\, \big(1 - \mathrm{SSIM}(\hat{I}_{\mathcal{G}}, I_{\text{sup}})\big),
\end{equation}
where $\lambda_{\text{ds}}$ balances high-frequency per-pixel accuracy with structural consistency.

To supervise 3D background integration, we introduce two additional regularizers applied to the background Gaussians $\mathcal{G}_{\text{bg}}$ obtained from densification of $\mathcal{G}_{\text{seed}}$. First, a radial shell loss penalizes primitives that leave a spherical shell:
\begin{equation}
    \mathcal{L}_{\text{shell}} = \frac{1}{|\mathcal{G}_{\text{bg}}|} \sum_{g_j \in \mathcal{G}_{\text{bg}}} \left( \max(0,\, d_j - R) + \max(0,\, \gamma R - d_j) \right)^2,
\end{equation}
where $d_j = \|\boldsymbol{\mu}_j - \mathbf{e}\|_2$. Second, a planarity loss discourages spiked Gaussians by aligning shortest axis $\mathbf{a}_j^{\min}$ with the radial direction $\hat{\mathbf{r}}_j = (\boldsymbol{\mu}_j - \mathbf{e}) / \|\boldsymbol{\mu}_j - \mathbf{e}\|_2$ from the shell centre:
\begin{equation}
    \mathcal{L}_{\text{plan}} = \frac{1}{|\mathcal{G}_{\text{bg}}|} \sum_{g_j \in \mathcal{G}_{\text{bg}}} \frac{s_j^{\max}}{s_j^{\min} + \epsilon} \Big(1 - \big| \mathbf{a}_j^{\min}\cdot \hat{\mathbf{r}}_j \big|\Big),
\end{equation}



During optimization, Gaussians from $\mathcal{G}_{\text{seed}}$ are densified into the final background $\mathcal{G}_{\text{bg}}$. While the primitives from $\mathcal{G}_{\text{seed}}$ are protected from pruning and deletion, this restriction does not apply to newly created Gaussians, allowing the model to capture fine details while removing redundant primitives. Additionally, scale-based culling is disabled across all background primitives to maintain extensive spatial coverage. The resulting unified representation combines fine-tuned scene and optimized background $\mathcal{G} = \mathcal{G}' \cup \mathcal{G}_{\text{bg}}$.

\section{Ablation: Generative Environment Completion}
\label{sec:ablation}

We outpainted the panorama using a generative model and used the same completed panorama in both VR conditions. Because the input represents a 360\textdegree environment, the missing content must be completed while preserving continuity across the spherical image. Several tested methods produced visible problems, for instance, \texttt{512-inpainting-ema} \cite{Rombach_2022_CVPR} and Nano Banana \cite{googlenanobanana2025} produced seams at the image boundaries, FLUX.1-Fill-dev~\cite{fluxfilldev2024} repeated parts of the scene, and FLUX.2-dev~\cite{flux2dev2025} produced color mismatches (see Fig.~\ref{fig:ablation-gen}).

Among the tested methods, FLUX.2-klein-base-9B with a 360\textdegree ERP LoRA provided the best overall result, with better color consistency and reduced seam artifacts (see Fig.~\ref{fig:ablation-gen}). The full candidate list and the two-stage wrap procedure used to remove the remaining ERP join are reported in the supplementary material.

\begin{figure*}[t]
  \centering
  \includegraphics[width=\textwidth]{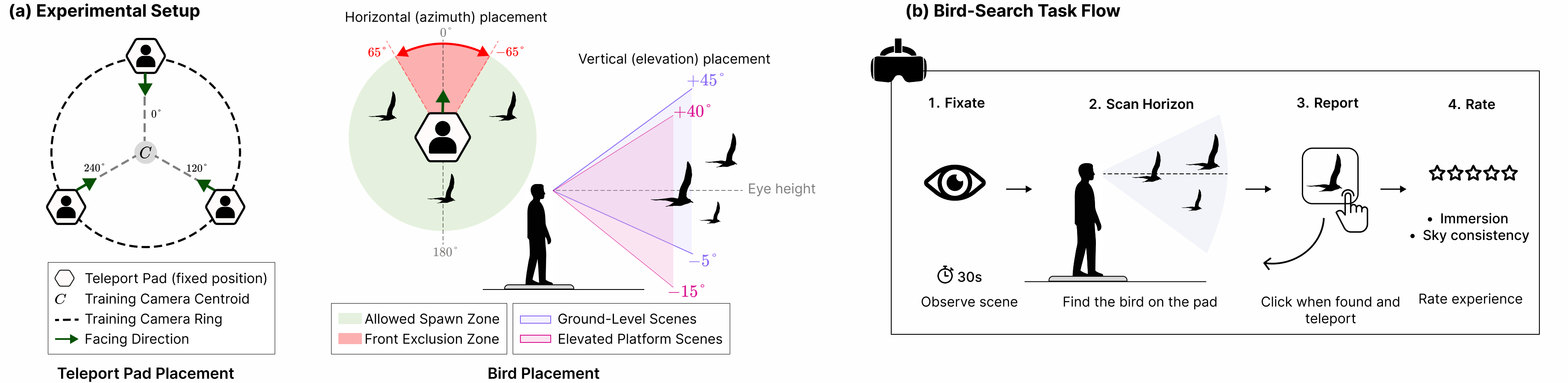}
  \caption{Bird-search task design. \textbf{(a)} Teleport pad and target
  placement logic: pads are positioned along the original camera
  trajectory, targets spawn outside a $\pm65^\circ$ front exclusion zone
  and within scene-dependent vertical elevation bands, with radial
  distance scaled to local scene geometry. \textbf{(b)} Task flow: the
  participant teleports to a pad, the corresponding target becomes
  visible, and the participant turns to locate and acquire it before
  proceeding to the next pad.}
  \label{fig:study-setup}
\end{figure*}

\section{User Study}
\label{sec:study}
We evaluated how background integration affects the perception of completed 3DGS scenes. The study follows a fully counterbalanced within-subjects design. Every participant experienced all four 3DGS scenes under both background rendering technologies (\our{} and Skybox), resulting in a repeated-measures evaluation. To control for potential ordering, learning, and fatigue effects, we counterbalanced the presentation order on two levels, i.e. the sequence of technologies was balanced between participants (forming two groups: Skybox-first vs. \our{}-first), while the presentation order of the four scenes within each block was randomized across participants using six distinct Latin-square sequences. The relevant ethics committee approved the user study. 

\subsection{Participants}
\label{sec:exp-participants}
We recruited $N = 36$ participants, satisfying an \emph{a priori} power analysis targeting ${\approx}80\%$ statistical power to detect a medium within-subjects effect size ($d_z \approx 0.5$, analytic minimum $N = 34$). 

Volunteers (31 male, 5 female) aged 20-37 years ($M = 26.2$, $SD = 3.8$) completed the study. Prior VR/AR experience was generally low (median 0\,years, $M = 0.8$, $SD = 1.3$), while self-reported gaming experience averaged 11.7\,years ($SD = 7.2$). Participants were divided into two groups. The only difference was the order in which we presented the methods. Between-group statistical tests confirmed that demographic variables did not differ significantly between the two order groups (all $p > 0.22$), confirming balanced baseline characteristics across groups (see Tab.~\ref{tab:participants}).

\subsection{Apparatus}
The VR environment was facilitated with a Meta Quest Pro VR HMD connected wirelessly to a dedicated computer through Meta Quest Air Link over a 5/6 GHz Wi-Fi network. The environments were rendered in Unity operating at a stable 90\,Hz refresh rate. The host rendering workstation was equipped with an Intel Core i7-14700K CPU (3.40\,GHz), 32.0\,GB of RAM, and an NVIDIA GeForce RTX~4070~SUPER GPU, running Windows~11 Home. Participants interacted with the scenes using Meta Quest Touch Pro hand-held controllers.

\subsection{Experimental Task}
\label{sec:exp-task}
\begin{figure}[t]
    \centering
    \includegraphics[width=\linewidth]{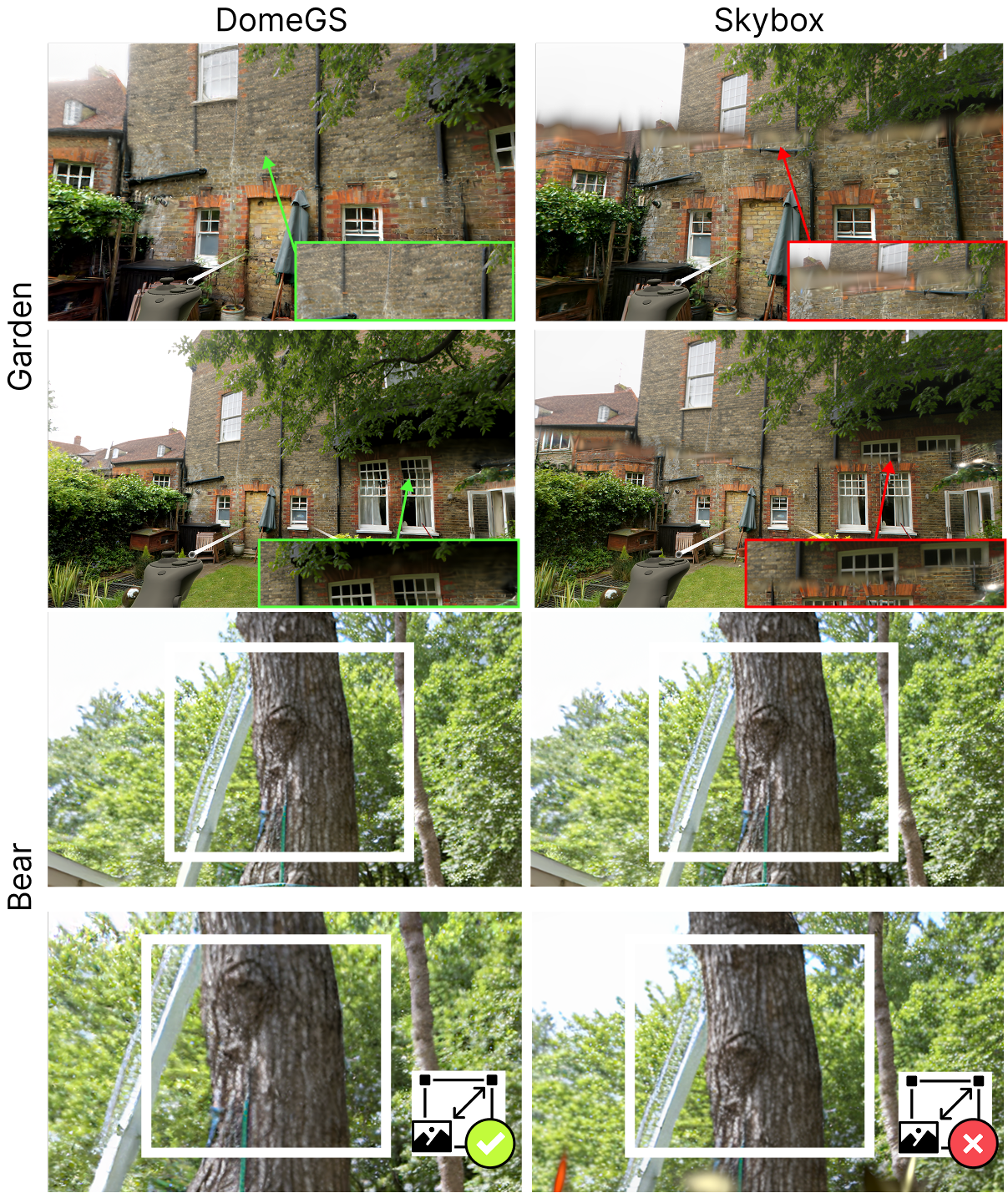}
    \caption{Parallax behavior under forward translation for \our{} (\emph{left}) and Skybox (\emph{right}), shown before and after movement
in two scenes. \emph{Top (Garden):} the Skybox's static background causes a visible mismatch with the foreground as the user moves, while \our{} maintains alignment. \emph{Bottom (Bear):} the Skybox tree fails to scale with proximity, whereas \our{}'s parallax reflects the change in distance.}
    \label{fig:parallax_comparison}
\vspace{-3mm}
\end{figure}

We designed a structured \textit{bird-search task} (see Fig.~\ref{fig:study-setup}) to elicit broad $360^\circ$ visual exploration while actively guiding participants’ attention to the upper hemisphere and horizon i.e. the area where the visual differences between the Skybox and our \our{} parallax are most prevalent (see Fig.~\ref{fig:parallax_comparison}). We deliberately selected such focused task instead of a free-exploration, exhaustive task to standardize visual attention across conditions and keep individual exposure within a concise time frame ($\approx$2–3 minutes per scene). 


\begin{table}[t]
\centering
\small
\setlength{\tabcolsep}{1.5pt}
\caption{Participant demographics by \textit{presentation order} (first \textit{Skybox}, or first \textit{\our{}}).}
\label{tab:participants}
\begin{tabularx}{\linewidth}{@{}l *{3}{>{\centering\arraybackslash}X}@{}}
\toprule
& \textbf{$N{=}36$ Participants} & \textbf{Skybox$\rightarrow$\our{}} & \textbf{\our{}$\rightarrow$Skybox} \\
\midrule
Age (yr)    & $M=26.2$ ($SD=3.8$) & $M=25.4$ ($SD=3.6$) & $M=26.9$ ($SD=4.0$) \\
VR/AR (yr)  & $M=0.8$ ($SD=1.3$)  & $M=1.1$ ($SD=1.7$)  & $M=0.4$ ($SD=0.6$)  \\
Gaming (yr) & $M=11.7$ ($SD=7.2$) & $M=10.9$ ($SD=5.8$) & $M=12.4$ ($SD=8.4$) \\
Gender (M/F) & 31 / 5     & 16 / 2     & 15 / 3 \\
\bottomrule
\end{tabularx}
\vspace{-3mm}
\end{table}

We placed teleport pads at viewpoints derived from the camera poses used to train the 3D Gaussian Splatting reconstruction, so that participants always view the scene from angles where the reconstruction is most reliable. This keeps viewpoint selection consistent across scenes, and helps ensure that immersion ratings reflect differences between the sky representations, rather than artifacts from viewing undersampled or poorly reconstructed scene components. We reused the same pad layout and visit order for both \our{} and Skybox within a scene, so participants followed an identical path regardless of which representation they experienced.

We used a procedural target placement system (see Fig.~\ref{fig:study-setup}), i.e. in each trial exactly one target (a bird) was assigned to each teleport pad and became visible only after the participant teleported to that pad. To avoid immediate target discovery upon arrival, we excluded targets from the frontal $\pm65^\circ$ region, requiring participants to search laterally and toward the rear. In addition, target elevation was calibrated to the scene geometry: $[-5^\circ,+45^\circ]$ for ground-level scenes (\emph{Garden}, \emph{Bear}) and $[-15^\circ,+40^\circ]$ for elevated platform scenes (\emph{Buddha}, \emph{Tower}). We adjusted these ranges to account for differences in scene geometry and keep search difficulty comparable across scene types. For example, in elevated scenes the greater amount of open sky makes targets easier to notice against a relatively uniform background, so  we shifted the elevation range downward and capped it at $+40^\circ$. In contrast, searches in the lower parts could be more difficult due to denser near-field geometry and greater occlusion. We therefore allowed a wider upward range, up to $+45^\circ$. 

Radial target distances were scaled to the environment geometry, ranging from $35\%$ to $85\%$ of the maximum inter-pad distance ($3.0$-$12.0\mathrm{m}$), with the participant's viewpoint fixed at a standard standing eye height of $1.6\mathrm{m}$ above the pad. To prevent spatial memorization across repeated visits, target positions alternated between two counterbalanced bird-spawn layouts ($\alpha$ and $\beta$), in which the teleport pads were fixed and only the
birds' azimuthal positions around each pad changed. 

\subsection{Independent Variables}
\label{sec:exp-iv}

The study follows a within-subjects $2 \times 2$ factorial design with two independent variables: \textit{technology} (i.e. \emph{Skybox} (baseline) vs. \emph{\our{}} (our)) and \textit{scene type} (i.e. \emph{Panoramic} vs. \emph{Close}). Every participant explores each 3DGS scene completed under both technologies, maintaining identical foreground Gaussians, the same panorama $I^\star$, and matched teleportation viewpoints (see Fig.~\ref{fig:study-setup}).


We selected four outdoor 3DGS reconstructions representing two spatial characteristics (see Fig.~\ref{fig:scenes}). The \emph{Panoramic} environments (\emph{Buddha}, \emph{Tower}) are open, aerial scenes dominated by distant backgrounds. In contrast, the \emph{Close} environments (\emph{Garden}~\cite{barron2022mip}, \emph{Bear}~\cite{instructnerf2023}) contain more nearby geometry and foreground objects. We included both types to ensure technology effects were not tied to a single capture style or source. Using different scene types also lets us examine whether the observed effects vary across reconstructions. Each participant experienced all four environments under both rendering conditions in a fully crossed repeated-measures design (4 scenes $\times$ 2 technologies). In the supplementary material \texttt{demo.mp4}, we also present examples of scenes using both technologies.

\subsection{Dependent Measures}
\label{sec:exp-dv}

We collected several types of measures, including in-task perceptual ratings, validated questionnaires, and behavioral data.

Following protocols demonstrating that single-item presence measures are psychometrically valid~\cite{bouchard2004reliability} and enable sensitive, low-intrusion in-VR assessment~\cite{bouchard2008anxiety, thorp2022temporal}, we adapted this approach using two single-item micro-probes inside the VR immediately after each \textit{bird-search} trial ($\approx$15-25\,s), rated on a 7-point Likert-like scale ($-3$ to $+3$). Higher \emph{IMM} scores indicate stronger perceived immersion and presence, while higher scores on the custom \emph{SKY} probe indicate better perceived background consistency and stability during movement and teleportation. This resulted in $N = 288$ observations per measure, allowing us to compare scenes while minimizing disruption to the VR experience. The exact item wording for \emph{IMM} and \emph{SKY} is reported in the supplementary material.

After exposure to either \our{} or Skybox condition, participants completed the \textit{Igroup Presence Questionnaire} (IPQ-PL) adapted for Polish samples ~\cite{strojny2014immersion}, previously used in the VR context \cite{11517219, Tadeja_Dataset_Generation_2021, Tadeja_3D_Photogrammetry_VR}, and adapted to their native tongue, comprising 13 items rated on a $-3$ to $+3$ scale. Following the IPQ-PL scoring procedure, we analyzed general presence (\emph{G1}), spatial presence (\emph{SP}), involvement (\emph{INV}), and experienced realism (\emph{REAL}), together with an overall IPQ score. At the end of the session, participants indicated their forced-choice or tied preference across four dimensions: (1) overall immersion, (2) close-range inspection, (3) panoramic viewing, and (4) willingness for prolonged stay. We also administered the \textit{Simulator Sickness Questionnaire} (SSQ) \cite{kennedy1993simulator} before and after the experiment to monitor potential cybersickness symptoms.

To verify that task difficulty remained balanced across background rendering conditions, we logged: (1) total \textit{bird-search} duration, defined as the time elapsed from search initiation until the final target was located; (2) head trajectory length, defined as cumulative head displacement excluding discrete teleportation jumps ($> 0.5$ m); and (3) target discovery latency, defined as the time required to locate individual target pads within each trial.


\subsection{Hypotheses}
\label{sec:exp-hypotheses}
While the Skybox baseline leaves the background statically locked to the viewpoint during locomotion, \our{} embeds the completed environment into finite 3D Gaussians. In VR, this provides natural motion parallax between foreground and background content during movement and teleportation (see Fig.~\ref{fig:parallax_comparison}). Therefore, we hypothesize improvement in perceived immersion, perceived visual consistency, and user preference:

\textbf{$H_1$} (\textbf{Immersion \& Consistency)}: \our{} yields significantly higher perceived immersion (\emph{IMM}) and background consistency under motion (\emph{SKY}) than Skybox.

\textbf{$H_2$} (\textbf{Presence \& Preference}): \our{} leads to higher perceived presence and is subjectively preferred over Skybox.

Furthermore, the accompanying experimental task was designed not to be difficult but to elicit broad visual exploration of the whole virtual environment. Therefore, we also hypothesize equivalence in the execution performance:

\textbf{$H_3$} (\textbf{Task Performance Equivalence}): Measurable behavior (i.e. search duration, head path length, and rotational exploration) does not differ between Skybox and \our{}.

\subsection{Experimental Protocol}
\label{sec:exp-protocol}

Individual sessions lasted $\approx$30-40 \, min (including $\approx$15-25\, min of total HMD exposure) and were carried out in five phases.
First, participants reviewed data protection, signed the consent form, completed a demographic survey, and filled out the baseline Simulator Sickness Questionnaire ($\text{SSQ}_{\text{pre}}$). Next, we familiarized themselves with the VR environment and interaction, including teleportation locomotion and target interaction, and the \textit{bird-search} task.

During the experimental phase, participants visited the four scenes (\emph{Buddha}, \emph{Tower}, \emph{Garden}, \emph{Bear}) in their assigned counterbalanced sequence under their initial representation condition (\our{} or Skybox) and target layout ($\alpha$ or $\beta$). Each scene followed a standard trial pipeline, i.e. a 30\,s observation phase on $\text{pad}_0$, the three-pad target search task, and an in-VR micro-probe ($\approx$15-25\,s). After the fourth scene, participants completed the mid-session sickness survey ($\text{SSQ}_{\text{mid}}$) and the first presence questionnaire ($\text{IPQ}_1$) during a mandatory $\approx$3\,min rest break.
Next, participants revisited the scenes in the same order under the alternate representation and mirrored target layout.

After exiting VR, participants completed the second presence questionnaire ($\text{IPQ}_2$), the forced-choice technology preference survey, and the final simulator sickness questionnaire ($\text{SSQ}_{\text{post}}$).


\subsection{Statistical Analysis}
\label{sec:exp-analysis}
We analyzed the micro-probe ratings, i.e. \emph{IMM} and \emph{SKY} ratings ($N = 288$) using Linear Mixed-Effects Models (LMMs; Gaussian family, \texttt{bobyqa} optimizer) fitted with Residual Maximum Likelihood (REML). Each model included used technology (\our{} vs.\ Skybox), scene (\emph{Buddha}, \emph{Tower}, \emph{Garden}, \emph{Bear}), and their two-way interaction as fixed effects, alongside a random intercept for participants.
We tested fixed effects with Satterthwaite's approximation for degrees of freedom and computed 95\% Wald confidence intervals. We verified residual normality using Q-Q plots. We adjusted post-hoc pairwise contrasts using the Holm-Bonferroni method. If the interaction was significant, we used the GAMLj\footnote{GAMLj: \url{https://gamlj.github.io/}} pairwise cell table (Holm over all 28 pairs) for follow-up.

For block-level IPQ subscales, we first tested pairwise differences ($\text{\our{}} - \text{Skybox}$) for normality using the Shapiro-Wilk test. If the assumption of normality was met ($p \ge 0.05$), paired $t$-tests were conducted; otherwise, two-tailed paired Wilcoxon signed-rank tests were used. $P$-values across the five IPQ subscales (\emph{G1}, \emph{SP}, \emph{INV}, \emph{REAL}, and \emph{Total}) were adjusted using the Holm method.

We recorded the presentation order of \our{} and Skybox to control for order effects, then encoded each preference response as $+1$ for (\our{}), $0$ for equal, and $-1$ for Skybox. Next, we evaluated directional preference using two-tailed exact binomial tests on non-tied responses against an equal-preference baseline ($p_0 = 0.5$). Moreover, we compared pre-session and post-session SSQ scores using paired tests to assess cybersickness onset.

To verify that search difficulty was unconfounded across rendering conditions, behavioral logs (i.e. search duration, head path, and yaw$+$pitch) were tested for equivalence with Two One-Sided Tests
(TOST) \cite{lakens2017equivalence} against a medium SESOI ($d_z{\approx}0.5$) mapped to raw bounds ($\pm 60$\,s, $\pm 6$\,m, $\pm 2600^\circ$). We also reported paired $t$-tests of a zero difference. As a sensitivity check that the same metrics can detect known difficulty, we compared \textit{Close} vs.\ \textit{Panoramic} person means collapsed across technology with paired $t$-tests. As a protocol check, we tested whether the person-level \our{}$-$Skybox difference itself differed by order in which participants experienced each completion method (Welch $t$ on the two order arms).

\section{User Study Results}
\label{sec:results}

\subsection{Immersion and Background Consistency ($H_1$)}
\label{sec:results-microprobes}


\begin{table}[t]
\centering
\small
\caption{LMM results for \emph{IMM} and \emph{SKY} micro-probe items ($N{=}288$ visits, 36 participants). $\Delta$ is the estimated marginal difference between \our{} and Skybox (\our{}${}-{}$Skybox) on the $-3\ldots{+}3$ scale; positive values indicate higher ratings for \our{}. 95\% CIs show the uncertainty around $\Delta$. Holm-adjusted $p$ values account for the two overall technology comparisons (SKY and IMM). $d$ is the corresponding standardized effect size. $technology\times scene$ (SKY $F(3,245){=}4.10$, $p{=}0.007$;
IMM $F(3,245){=}0.81$, $p{=}0.488$) tested whether the technology difference varies across scenes.}
\label{tab:microprobe_omnibus}
\smallskip
\begin{tabular}{@{}lccccc@{}}
\toprule
Item & $\Delta$ & 95\% CI & $F(1,245)$ & Holm $p$ & $d$ \\
\midrule
SKY & $+1.24$ & $[0.95, 1.53]$ & $70.70$ & $<0.001$ & $0.85$ \\
IMM & $+0.44$ & $[0.21, 0.68]$ & $14.01$ & $<0.001$ & $0.35$ \\
\bottomrule
\end{tabular}\\[0.4em]
\end{table}

\begin{figure}[t]
\centering
\includegraphics[width=\linewidth]{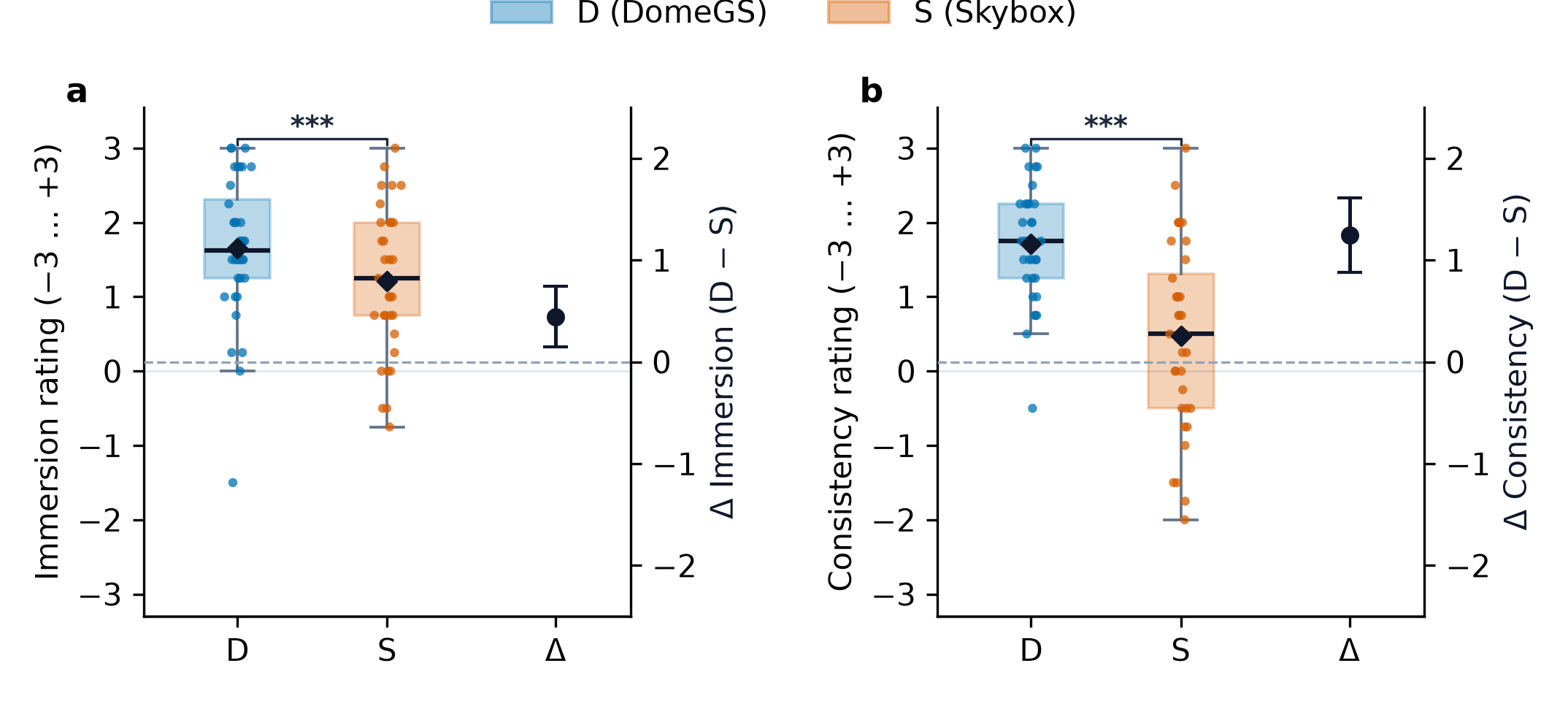}
\caption{Participant-level micro-probe estimation plots ($N = 36$): (a)~Perceived Immersion (\emph{IMM}) and (b)~Background Consistency under motion (\emph{SKY}). Left subpanels show raw distributions for (B) \our{} and (S) Skybox via boxplots (medians and IQRs), jittered points (individual participants), and diamond markers (group means). Right subpanels display the paired mean difference ($\text{\our{}} - \text{Skybox}$) with its 95\% confidence interval (dashed line indicates zero difference; $^{*}$ Holm $p < .05$, $^{***}$ Holm $p < .001$).}
\label{fig:microprobe_results}
\vspace{-3mm}
\end{figure}

In terms of \textbf{perceived immersion} (\emph{IMM}), the immersion ratings were significantly higher for \our{} compared to Skybox (see Fig.~\ref{fig:microprobe_results}a; $\Delta{=}{+}0.44$, 95\% CI $[0.21, 0.68]$, Holm $p{<}0.001$, $d{=}0.35$). The \textit{technology} $\times$ \textit{scene} interaction was not significant ($F(3,245){=}0.81$, $p{=}0.488$), indicating that the advantage of \our{} over the baseline did not significantly differ across scenes.

With respect to \textbf{background consistency under motion} (\emph{SKY}), \our{} ratings were significantly higher than Skybox on average (see Fig.~\ref{fig:microprobe_results}b; $\Delta{=}{+}1.24$, Holm $p<0.001$, $d{=}0.85$). The \textit{technology} $\times$ \textit{scene} interaction was significant ($F(3,245){=}4.10$, $p{=}.007$), indicating that although $\our{}$ was overall better than $Skybox$, the size of this advantage varied across scenes. Scene-wise contrasts showed that \our{} significantly outperformed the baseline in three scenes, with the largest difference in \emph{Buddha} ($\Delta{=}+2.08$, Holm $p<0.001$), followed by \emph{Bear} ($\Delta{=}{+}1.14$, Holm $p{=}0.002$), \emph{Garden} ($\Delta{=}+1.08$, $p{=}0.004$). The difference for the \emph{Tower} scene was smaller ($\Delta=+0.67$) and was not statistically significant after correction (Holm $p=0.225$). The \emph{Tower} also had relatively high ratings under the Skybox condition, compared with the remaining three scenes (all Holm $p{<}0.001$). 

The LMM results used to evaluate $H_1$ are summarized in Tab.~\ref{tab:microprobe_omnibus} and Fig.~\ref{fig:microprobe_results}. Full model coefficients, random-effect estimates, and scene-wise pairwise comparisons are provided in the supplementary material. Based on these results, we conclude that \textbf{$H_1$} is supported and that \our{} yields significantly higher perceived immersion (\emph{IMM}) and background consistency under motion (\emph{SKY}) than the traditional Skybox baseline.

\subsection{Preconceived Presence and Preference ($H_2$)}
\label{sec:results-preference}

\begin{figure}[t]
\centering
\includegraphics[width=\linewidth]{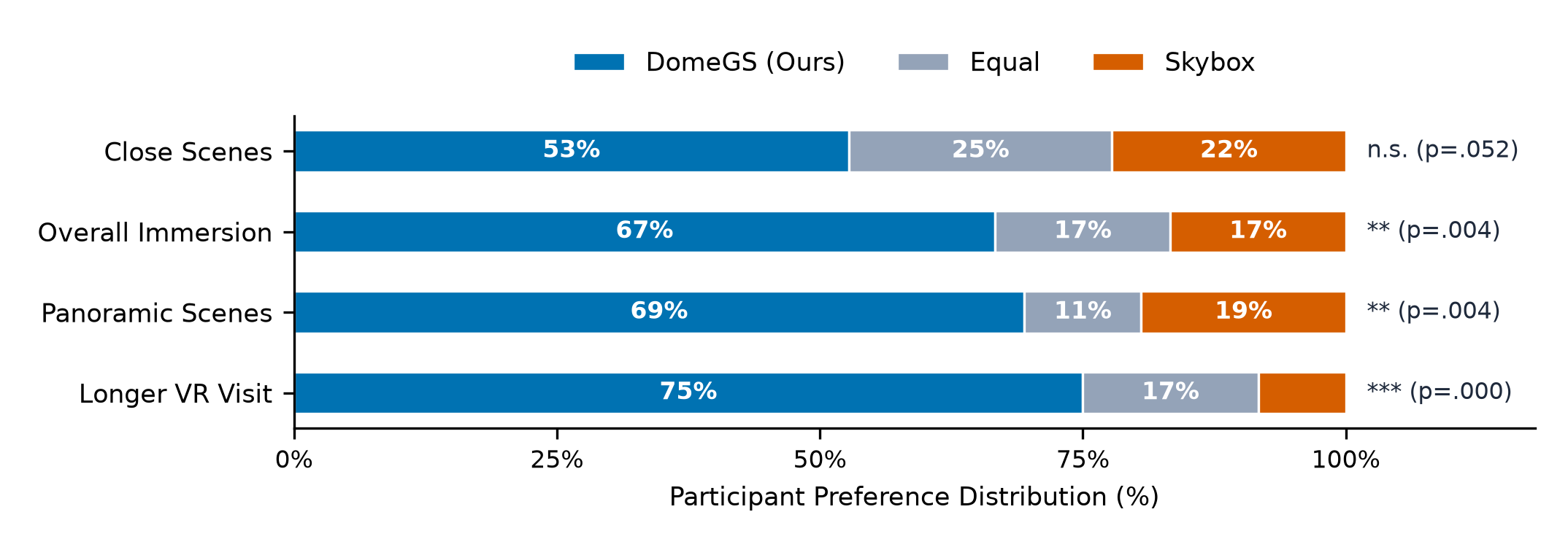}
\caption{End-of-session comparative preference distributions ($N = 36$). Stacked horizontal bars show the proportion of participants favoring \our{} (blue), Equal (gray), and Skybox (orange). Statistical annotations denote Holm-adjusted two-tailed exact binomial tests against chance on non-tied choices ($^{*}p_{\text{adj}} < 0.05$, $^{**}p_{\text{adj}} < 0.01$, $^{***}p_{\text{adj}} < 0.001$).}
\label{fig:preference_distribution}
\vspace{-3mm}
\end{figure}

Participants generally preferred \our{} over Skybox for \textbf{overall immersion, panoramic viewing, and willingness to prolong their stay within the virtual scenes} (see Fig. \ref{fig:preference_distribution}). The strongest preference was for longer VR visit, with $75.0\%$ of participants choosing \our{} compared with only $8.3\%$ for Skybox (Holm $p<0.001$). For \textit{Panoramic} scenes and overall immersion, \our{} was chosen by $69.4\%$ (Holm $p=0.004$) and $66.7\%$ (Holm $p=0.004$) of participants, respectively.

In \textit{Close} scenes (i.e. \emph{Garden}, \emph{Bear}), \our{} was preferred by $52.8\%$ of participants, compared with $22.2\%$ for Skybox, with the difference not reaching the corrected significance threshold (Holm $p=0.0522$, $25.0\%$ of participants reported no preference, the highest tie rate among the four measures). This difference across scene types is consistent with theories of visual depth perception and spatial layout \cite{cutting1995perceiving, howard2002seeing}. In cluttered close-range scenes, dense foreground geometry already provides sufficient local depth and motion cues, which may reduce reliance on distant background cues. In contrast, open panoramic scenes are dominated by the horizon and distant environment, making the spatial continuity offered by our \our{} important for maintaining immersion \cite{cutting1995perceiving, howard2002seeing}.

At the same time, preference did not differ by order in which participants experienced the \textit{technology} (Fisher’s exact test on non-tied responses, all $p{>}0.58$). Among participants with complete, non-tied preference profiles, a Cochran’s test likewise found no significant differences in \our{}-choice rates across the four scenes.

IPQ scores showed acceptable internal consistency in both conditions (\our{} $\alpha{=}0.86$; Skybox $\alpha{=}0.80$). Shapiro-Wilk tests on paired differences were inconsistent across subscales (some $p{<}0.05$, others $p{\ge}0.05$), therefore we applied two-sided Wilcoxon signed-rank tests to all five outcomes and Holm-adjusted $p$-values. 

\begin{figure*}[t]
\centering
\includegraphics[width=0.9\textwidth]{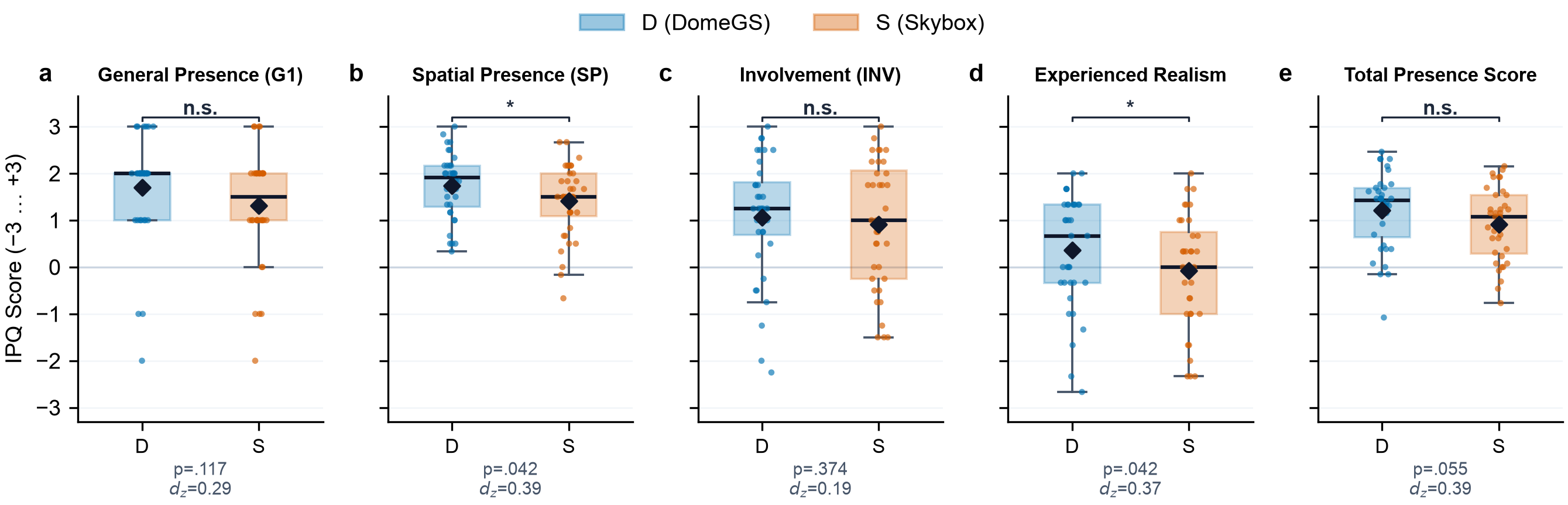}
\caption{Block-level IPQ presence scores across representation conditions ($N = 36$): (a)~general presence (\emph{G1}), (b)~spatial presence (\emph{SP}), (c)~involvement (\emph{INV}), (d)~experienced realism (\emph{REAL}), and (e)~total presence. Box plots show medians and interquartile ranges (IQRs); jittered points show individual participant ratings on the $[-3, +3]$ scale (\our{} in blue, Skybox in orange); and diamond markers indicate group means. Overhead brackets denote Holm-adjusted significance ($^{*}$Holm $p < 0.05$, n.s.: Holm $p \ge 0.05$).}
\label{fig:ipq_subscales}
\end{figure*}

\our{} received higher mean scores than Skybox across all IPQ subscales (see Fig.~\ref{fig:ipq_subscales}). After Holm correction, spatial presence (\emph{SP}) ($1.73$ vs.\ $1.41$, $\Delta=+0.32$, Holm $p=0.042$) and experienced realism (\emph{REAL}) ($0.36$ vs.\ $-0.07$, $\Delta=+0.44$, Holm $p=0.042$) were significant.  General presence (\emph{G1}), involvement (\emph{INV}), and total score favored \our{} but were not significant after correction (Holm $p=0.117$, $p=0.374$, and $p=0.055$, respectively). The full Wilcoxon statistics, including means, confidence intervals, and test statistics for all subscales, are reported in the supplementary material. The interaction between presentation order and technology exposure was not statistically significant for any measure (all Holm $p \geq 0.142$), suggesting that participants' preferences were not significantly influenced by order of exposure to each technology.

These findings provide partial support for $H_2$, suggesting that \our{} leads to statistically higher spatial presence (\emph{SP}) ($1.73$ vs.\ $1.41$, $\Delta=+0.32$, Holm $p=0.042$) and experienced realism (\emph{REAL}) ($0.36$ vs.\ $-0.07$, $\Delta=+0.44$, Holm $p=0.042$), and is significantly preferred over Skybox for \textit{Panoramic} scenes.

\subsection{Task Performance and Manipulation Checks ($H_3$)}
\label{sec:results-task}
\begin{figure}[t]
\centering
\includegraphics[width=\linewidth]{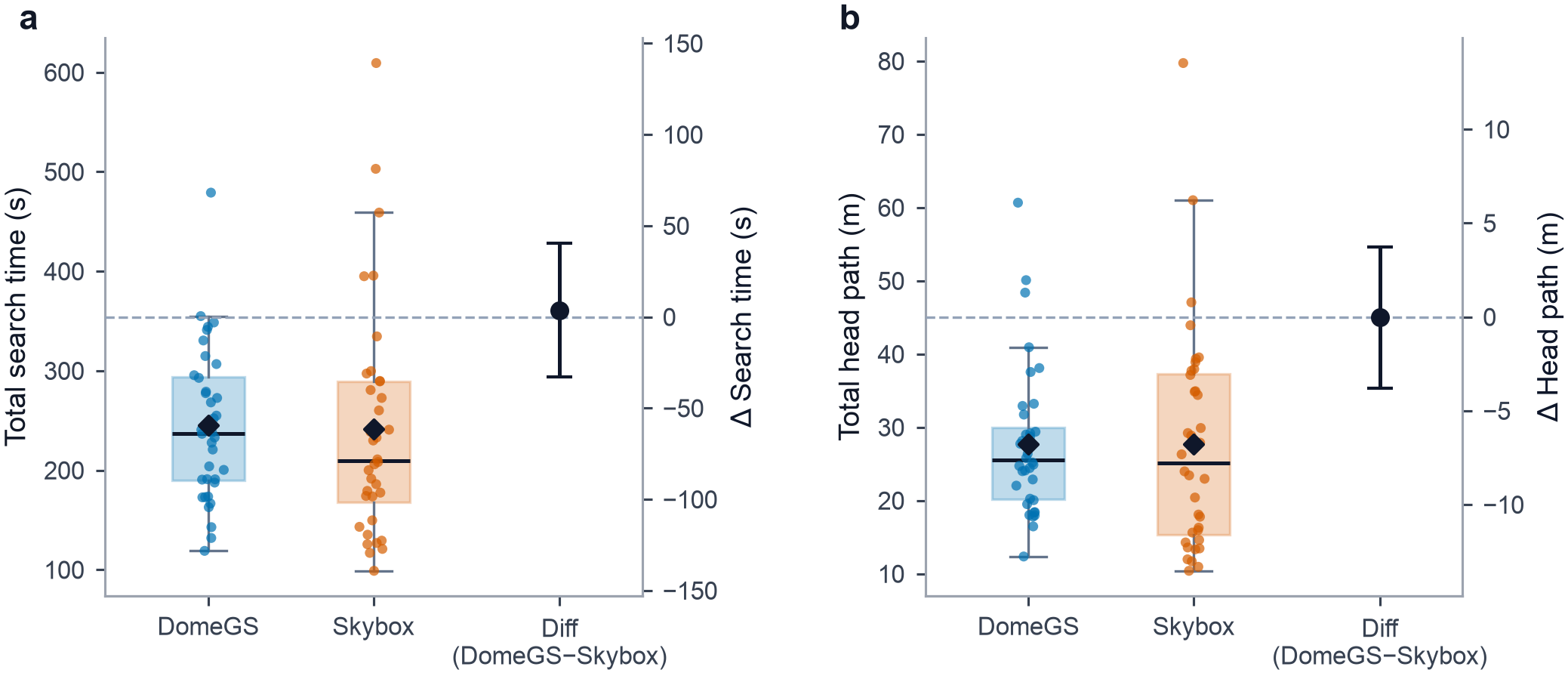}
\caption{Participant-level task performance estimation plots, i.e. (a)~total search time and (b)~total head path length across all scenes. Left subpanels show raw distributions for \our{} and Skybox via boxplots (medians and IQRs), jittered points (individual participants), and diamond markers (group means). Right subpanels display the paired mean difference ($\text{\our{}-\text{Skybox}}$) with its 95\% confidence interval (dashed line indicates zero difference; n.s. for Holm $p \ge 0.05$).}
\label{fig:task_estimation}
\vspace{-3mm}
\end{figure}

To evaluate $H_3$ whether measurable behavior did not differ between Skybox and \our{}, we tested whether person-level search duration, head path, and rotational exploration (yaw$+$pitch) were equivalent under \our{} vs.\ Skybox (see Fig.~\ref{fig:task_estimation}, Tab.~\ref{tab:results}). TOST against a medium SESOI ($d_z{\approx}0.5$; $\delta{=}\pm 60$\,s, $\pm 6$\,m, $\pm 2600^\circ$) rejected nonequivalence for all three metrics (all $p_{\mathrm{TOST}}{\le}0.005$): each 90\% CI of $\Delta$ (\our{}$-$Skybox) lay inside its bound.

In addition, we analyzed head orientation (yaw $+$ pitch) to verify that participants actively engaged with the background. Participants showed broad $360^\circ$ visual exploration in both conditions, characterized by low mean resultant lengths (\our{}: $R=0.094$; Skybox: $R=0.113$; paired Holm $p=0.721$, $d_z=0.23$) and wide circular standard deviations ($128.7^\circ$ vs.\ $123.8^\circ$). They spent more than two-thirds of search time looking above the horizon toward the sky (pitch $>0^\circ$: $68.8\%$ vs.\ $68.4\%$; $\Delta=+0.44\%$, Holm $p=1.000$, $d_z=0.03$). Mean vertical pitch was also matched ($7.70^\circ$ vs.\ $7.30^\circ$, Holm $p=1.000$). Together, these results suggest that participants explored the background similarly in both conditions, with no evidence of a systematic gaze-angle bias.

Participants spent more time, followed longer paths, and scanned more in $Close$ scenes than in $Panoramic$ scenes (all $p < 0.01, |d_z| \ge 0.49$), indicating that these metrics were sensitive to scene layout and consistent with the greater movement and visual-search demands of $Close$ scenes. Interactions between presentation order and technology were not significant for either search time ($p = 0.135$) or path ($p = 0.127$). Gaze heatmaps, $Close$ vs.\ $Panoramic$ tables, order-group diagnostics, and the remaining Euler TOST tests are in the supplementary material.

Together, these findings support $H_3$ that behavioral load (i.e. search duration, head path, and rotational exploration) was equivalent across technologies within a medium effect-size bound, confirming matched task difficulty. Thus, we concluded that the accompanying experimental task was equally difficult across all conditions and elicited the same broad visual exploration pattern.

\begin{table}[t]
\centering
\small
\setlength{\tabcolsep}{4.5pt}
\caption{Task-load equivalence ($H_3$; $n{=}36$ paired), $\Delta$ is \our{}$-$Skybox with the TOST 90\% CI. Bounds $\delta$ implement a medium SESOI ($d_z{\approx}0.5$), $p_{\mathrm{TOST}}{=}\max(p_{\mathrm{lower}},p_{\mathrm{upper}})$.}
\label{tab:results}
\smallskip
\begin{tabular}{@{}lcccc@{}}
\toprule
Measure & $\delta$ & $\Delta$ [90\% CI] & NHST $p$ & $p_{\mathrm{TOST}}$ \\
\midrule
Search time (s) & $\pm 60$ & $-6.7$ [$-38.8$, $25.4$] & $.727$ & $.004$ \\
Head path (m) & $\pm 6$ & $-0.40$ [$-3.74$, $2.95$] & $.842$ & $.004$ \\
Yaw$+$pitch ($^\circ$) & $\pm 2600$ & $-220$ [$-1694$, $1254$] & $.803$ & $.005$ \\
\bottomrule
\end{tabular}
\end{table}

\subsection{Cybersickness and Comfort}
\label{sec:results-ssq}
SSQ scores~\cite{kennedy1993simulator} analyzed with Wilcoxon did not differ between $\our{}$ and $Skybox$; Wilcoxon throughout, Shapiro-Wilk $p{\le}0.005$). Total was $16.5$ vs.\ $17.9$ (\our{} vs.\ Skybox;
$\Delta{=}-1.35$, 95\% CI $[{-}5.46,\,2.75]$,
$W{=}117$, Holm $p{=}1.000$, $d_z{=}-0.11$);
Nausea, Oculomotor, and Disorientation were likewise n.s.\ (all Holm $p{=}1.000$). Scores stayed in the mild range (Total ${<}20$). Pre-to-post Total accumulation was not significant ($14.4$ to $18.0$, $t(35){=}1.21$, Holm $p{=}0.931$); Disorientation rose modestly ($7.3$ to $16.2$, Wilcoxon Holm $p{=}0.05$, $d_z{=}0.43$). The full SSQ descriptives, paired technology contrasts and pre-to-post accumulation are reported
in the supplementary material.


\subsection{Automated Vision-Language Quality Evaluation}
As an additional sanity check, alongside human perceptual evaluations, we conducted an objective visual quality assessment using a state-of-the-art vision-language model (Qwen3-VL-32B-Instruct) as an automated visual inspector across all test scenes. The rendered viewpoints were evaluated using two protocols:

(1) \textbf{Absolute Quality Scoring} (1-5 scale), where renders were evaluated across three visual criteria: Boundary Continuity (B), Lighting Coherence (C), and Perspective/Horizon Alignment (P).
    
(2) \textbf{Pairwise Comparison} (-2 to +2 scale), where renders from \our{} and the Skybox were evaluated head-to-head with position swapping. Positive scores indicate preference for \our{}.

As summarized in Tab.~\ref{tab:vlm_evaluation}, \our{} consistently outperforms the traditional Skybox baseline across all datasets and metrics. In absolute scoring, \our{} achieves higher scores in boundary transition, lighting coherence, and perspective alignment. Furthermore, the position-swapped pairwise evaluation confirms a strong preference for \our{} across all scenes. 

\begin{table}[t]
\centering
\small
\setlength{\tabcolsep}{4pt}
\caption{Automated visual quality evaluation using Qwen3-VL-32B. Absolute Scores (1-5 scale): Boundary Continuity (B), Lighting Coherence (C), and Perspective Alignment (P). Pairwise Preference ($-2$ to $+2$ scale): Relative comparison with position swapping ($+2 = \text{\our{}}$ win, $-2 = \text{Skybox}$ win).}
\label{tab:vlm_evaluation}
\begin{tabularx}{\linewidth}{l c >{\centering\arraybackslash}X >{\centering\arraybackslash}X c}
\toprule
& & \textbf{Skybox (Baseline)} & \textbf{\our{} (Ours)} & \\
\textbf{Scene} & \textbf{Typology} & \textbf{B / C / P} & \textbf{B / C / P} & \textbf{Pref. Score} \\
\midrule
Tower  & Panoramic & 3.22 / 3.38 / 3.31 & \textbf{4.00 / 4.10 / 3.71} & \textbf{+0.747} \\
Buddha & Panoramic & 2.36 / 2.90 / 2.79 & \textbf{3.43 / 3.58 / 3.58} & \textbf{+1.083} \\
Garden & Close     & 2.72 / 2.96 / 2.87 & \textbf{3.69 / 3.68 / 3.70} & \textbf{+0.475} \\
Bear   & Close     & 2.54 / 2.69 / 2.67 & \textbf{3.38 / 3.46 / 3.52} & \textbf{+0.383} \\
\midrule
Mean   & -         & 2.71 / 2.98 / 2.91 & \textbf{3.63 / 3.71 / 3.63} & \textbf{+0.672} \\
\bottomrule
\end{tabularx}
\end{table}

\section{Discussion}

In our study, we compared two main conditions that differed only in the far-field representation, i.e. the traditional infinite skybox baseline provided no translational parallax as opposed to the proposed finite \our{} binding skydome with 3DGS. The potential improvements therefore reflect the far-field representation itself by unifying the near-field scene and the far-field skydome into a single 3DGS asset, ensuring that the entire environment responds cohesively to user movement. In contrast, prior 3DGS pipelines either suppress the sky~\cite{kulhanek2024wildgaussians, wang2025look} or generate entire panoramic worlds~\cite{yang2025layerpano3d, zhou2024dreamscene360}. These approaches change image content and representation together. 

Moreover, to evaluate our approach's quality, we conducted a controlled user study to isolate the effect of the far-field representation. The micro-probe ratings following low-intrusion presence assessment protocols \cite{bouchard2004reliability, bouchard2008anxiety, thorp2022temporal} suggest that the enhancement was largest for the items targeting the background consistency under motion (\emph{SKY}, $p{<}0.001$, $d{=}0.85$) and smaller but reliable for perceived immersion (\emph{IMM}, $p{<}0.001$, $d{=}0.35$, uniform across scenes), which is consistent with presence being shaped by multiple sensorimotor cues \cite{cummings2016immersive, slater2009place}. Furthermore, the significant $technology \times scene$ interaction for \emph{SKY} shows where this benefit was concentrated, namely, it was largest in the \emph{Buddha} scene and not significant in \emph{Tower}, where Skybox ratings were already high. Since both scenes belong to the same \emph{Panoramic} category, the simple taxonomy of panoramic and non-panoramic scenes alone cannot explain this difference. A possible explanation can be tentatively attributed to what the initial reconstruction missed in each scene, i.e. in the case of the \emph{Tower} scene, it was mainly sky, which was well-approximated with the traditional skybox's infinite-distance assumption. Whereas the \emph{Buddha} scene also includes closer terrain and rock formations that this assumption cannot represent correctly, but \our{} can result in a much larger improvement (Fig.~\ref{fig:scenes}).

We also saw a similar pattern among questionnaire results. Among the IPQ subscales, only the items directly related to the depth cues, i.e. spatial presence ($p=0.042$), as well as experienced realism ($p=0.042$), related to the scene behaving as a physical space should as the viewer moves, were statistically significant.

In terms of participants' declared preferences, the only preference contrast that did not reach significance was for \emph{Close} scenes ($p{=}0.052$), which also had the highest rate of tied responses, which may reflect depth-cue redundancy \cite{cutting1995perceiving, howard2002seeing}. In cluttered near-field scenes, nearby geometry already provides strong parallax and occlusion cues, so the background may contribute less additional information. While in open panoramic scenes, where the horizon dominates the view but parallax is naturally weak at that distance, background coherence, the absence of seams at the horizon, is likely the more salient cue driving
preference.

Moreover, these observed gains carried no added cost, with SSQ scores remaining mild and statistically non-significant between the Skybox and \our{} approaches. This is consistent with skydome parallax being geometrically compatible with the user's own translation, avoiding the visual-vestibular conflict contributing to the induction of simulator sickness symptoms \cite{laviola2000discussion}.

\section{Conclusion}
In this work, we offer \our{}, a new background rendering technique that provides enhanced visual fidelity for completing 3DGS scenes. Our pipeline is specifically designed to preserve depth cues and realism in an immersive VR environment. To achieve this, we integrated $360^\circ$-aware generative outpainting into a Gaussian-based skydome, enabling correctly aligned, coherent background completion without degrading the primary reconstructed scene.
    
We systematically compared our approach with a traditional skybox baseline via a comprehensive within-subjects user study with $N=36$ participants. The results of this experimental comparison demonstrate that 3DGS significantly improves background consistency under motion, spatial presence, and experienced realism compared to the classical skybox baseline.

\section*{Acknowledgments}
This work of P.~Spurek, A.~Pechko, was supported by the project ``Effective rendering of 3D objects using Gaussian Splatting in an Augmented Reality environment'' (FENG.02.02-IP.05-0114/23) carried out within the First Team programme of the Foundation for Polish Science co-financed by the European Union under the European Funds for Smart Economy 2021-2027 (FENG). The work of P.~Borycki was supported by the National Centre of Science (Poland) Grant No. 2025/57/N/ST6/04389. The work of J.~Szelewska was supported by the National Centre of Science (Poland) Grant No. 2025/57/N/ST6/04390.

The authors acknowledge the use of generative AI tools, namely GPT-5.6, Gemini Flash 3.6, and Grammarly, solely for editorial assistance, grammar enhancement, and improving the text flow. The figures use icons from flaticon \footnote{flaticon, \url{https://www.flaticon.com/}}. Due to limited space the exact author attributions can be found in supplementary materials.


\bibliographystyle{abbrv-doi}

\bibliography{template}
\end{document}